# K-Implementation

**Dov Monderer**                                          DOV@IE.TECHNION.AC.IL
**Moshe Tennenholtz**                                MOSHET@IE.TECHNION.AC.IL
*Faculty of Industrial Engineering and Management*
*Technion – Israel Institute of Technology*
*Haifa 32000, Israel*

## Abstract

This paper discusses an interested party who wishes to influence the behavior of agents in a game (multi-agent interaction), which is not under his control. The interested party cannot design a new game, cannot enforce agents' behavior, cannot enforce payments by the agents, and cannot prohibit strategies available to the agents. However, he can influence the outcome of the game by committing to non-negative monetary transfers for the different strategy profiles that may be selected by the agents. The interested party assumes that agents are rational in the commonly agreed sense that they do not use dominated strategies. Hence, a certain subset of outcomes is implemented in a given game if by adding non-negative payments, rational players will necessarily produce an outcome in this subset. Obviously, by making sufficiently big payments one can implement any desirable outcome. The question is what is the cost of implementation? In this paper we introduce the notion of $k$-implementation of a desired set of strategy profiles, where $k$ stands for the amount of payment that need to be actually made in order to implement desirable outcomes. A major point in $k$-implementation is that monetary offers need not necessarily materialize when following desired behaviors. We define and study $k$-implementation in the contexts of games with complete and incomplete information. In the latter case we mainly focus on the VCG games. Our setting is later extended to deal with mixed strategies using correlation devices. Together, the paper introduces and studies the implementation of desirable outcomes by a reliable party who cannot modify game rules (i.e. provide protocols), complementing previous work in mechanism design, while making it more applicable to many realistic CS settings.

## 1. Introduction

The design and analysis of interactions of self-interested parties are central to the theory and application of multi-agent systems. In particular, the theory of economic mechanism design or, more generally, implementation theory (Maskin, 1999; Maskin & Sjostrom, 2002) has become a standard tool for researchers in the areas of multi-agent systems and e-commerce (Rosenschein & Zlotkin, 1994; Nisan & Ronen, 1999; Shoham & Tennenholtz, 2001; Feigenbuam & S, 2002; Tennenholtz, 1999; Papadimitriou, 2001). In classical mechanism design[1] a center defines an interaction for self-motivated parties that will allow it to obtain some desired goal (such as maximizing revenue or social welfare) taking the agents' incentives into account. This perspective has been largely motivated by the view of the center as a government or a seller that can define and control the rules of interaction. However, in many distributed systems and multi-agent interactions, interested parties cannot control the rules of interactions. A network manager for example cannot simply change the communication protocols in a given distributed systems in order to lead to desired behaviors, and a broker cannot change the rules in which goods are sold by an agency auctioneer to the public. The focus of this paper is on how a reliable interested party, which cannot change the rules of interaction, and cannot enforce behavior, can obtain its desired goals (in service of the community or for its own benefits). The reliable party has only one source of power: its reliability. It can commit to payments to the different agents, when certain observable outcomes will be reached, and the agents can be sure that they will be paid appropriately.

---

1. See e.g., Fudenberg and Tirole (1991), Chapter 7, or Mas-Colell, Whinston, and Green (1995), Chapter 23.

In our work we introduce the study of implementation of desired behaviors by interested party as above.[2] There are two major issues that make the task non-trivial and challenging:

**1.** The interested party may wish to assume as little as possible about agents' rationality. Ideally, all that will be assumed is that an agent does not adopt a strategy if it is dominated by another strategy.

**2.** The interested party may wish to minimize its expenses.

Consider the following simple congestion setting.[3] Assume that there are two agents, 1 and 2, that have to select among two service providers (e.g., machines, communication lines, etc.) One of the service providers, $f$, is a fast one, while the other, $s$, is a slower one. We capture this by having an agent obtaining a payoff of 6 when he is the only one that uses $f$, and a payoff of 4 when he is the only one who uses $s$. If both agents select the same service provider then its speed of operation decreases by a factor of 2, leading to half the payoff. That is, if both agents use $f$ then each one of them obtains a payoff of 3, while if both agents use $s$ then each one of them obtains 2. In a matrix form, this game is described by the following bimatrix:

$$
M = \quad
\begin{array}{c|c|c|}
 & f & s \\
\hline
f & 3 \quad\quad 3 & 6 \quad\quad 4 \\
\hline
s & 4 \quad\quad 6 & 2 \quad\quad 2 \\
\hline
\end{array}
$$

---

Assume that our reliable interested party may wish to prevent the agents from using the same service provider (leading to low payoffs for both). Then it can do as follows: it can promise to pay agent 1 a value of 10 if both agents will use $f$, and promise to pay agent 2 a value of 10 if both agents will use $s$. These promises transform $M$ to the following game:

$$
M' = \begin{array}{c c}
 & \begin{array}{c c} f & \quad s \end{array} \\
\begin{array}{c} f \\ \\ s \end{array} &
\begin{array}{|c|c|}
\hline
13 \quad\quad & 6 \quad\quad \\
\quad\quad 3 & \quad\quad 4 \\
\hline
4 \quad\quad & 2 \quad\quad \\
\quad\quad 6 & \quad\quad 12 \\
\hline
\end{array}
\end{array}
$$

Notice that in $M'$, strategy $f$ is dominant for agent 1, and strategy $s$ is dominant for agent 2. As a result the only rational strategy profile is the one in which agent 1 chooses $f$ and agent 2 chooses $s$. Hence, the interested party implements one of the desired outcomes. Moreover, given that the strategy profile $(f, s)$ is selected the interested party will have to pay nothing. It has just implemented (in dominant strategies) a desired behavior (obtained in one of the Nash equilibria) with zero cost, relying only on its creditability, without modifying the rules of interactions or enforcing any type of behavior.

Similar simple examples can be found in other contexts (see e.g., Segal (1999), footnote 30, and Dybvig and Spatt (1983), Spiegler (2000)). Our work advocates the following line of thought. Instead of reasoning about how agents will behave in the given protocol, we may wish to cause agents to follow particular behaviors by making them desirable, using monetary offers. An important point is that the monetary offers need not necessarily be fully materialized when agents follow the desired behavior.

More formally, in this paper we introduce the notion and study of $k$-implementation of a desired set of strategy profiles, where $k$ stands for the amount of payment that need to be actually made in order to implement the desirable outcomes.[4] Section 3 provides a characterization of $k$-implementation of a single pure strategy profile for finite games and infinite regular games with complete information. This provides an effective algorithm for determining the optimal monetary offers to be made in order to implement a desired outcome, while minimizing expenses. In Section 4 we address the problem of finding a $k$-implementation of a set of strategy profiles. We show that the general problem in this regard is NP-hard, and consider a modification of $k$-implementation, titled exact implementation, under which the problem becomes tractable.[5]

---

4. Notice that this perspective is in the spirit of work on Artificial Social Systems in AI (see e.g., Shoham and Tennenholtz (1995)), where we search for some form of "modification" to the system, such that given the modified system, and assuming agents tend to work individually, a desirable outcome will be obtained.

Games with incomplete information introduce further challenges. In particular, in Section 5 we consider the VCG mechanisms for combinatorial auctions [6]. This setting has interesting characteristics since the interested party cannot in general see the agents' types and needs to decide on appropriate payment only based on observed behaviors. We show that in general 0-implementation (i.e. implementation with zero cost) in settings with incomplete information is impossible, but any ex-post equilibrium of a frugal VCG mechanism is 0-implementable.

In Section 6 we study the important case of mixed strategies. In that context, unless we assume algorithmic observability, the interested party can observe only the actions selected and not the probabilistic process leading to the selection, and therefore our earlier results do not apply. For example, consider the simple routing problem above, one may wish to consider the implementation of a more "fair" outcome, such as the one obtained in the mixed strategy Nash equilibrium of the game $M$. In order to address this issue, we introduce the concept of implementation devices, and show that any mixed strategy equilibrium is 0-implementable with an implementation device. We also show that any correlated equilibrium has this property.

## 2. k-implementation

A *pre-game* in strategic form is a pair $G = (N, X)$, where $N = \{1, 2, \cdots, n\}$ is the set of *players*, $X = X_1 \times X_2 \times \cdots X_n$, where for every $i$, $X_i$ is the set of strategies available to player $i$. Let $i$ be a player, the set of strategy profiles of all other players is denoted by $X_{-i}$, and a generic element in $X_{-i}$ is denoted by $x_{-i}$.

A *payoff function vector* is an $n$-tuple $U = (U_1, U_2, \cdots, U_n)$, where $U_i : X \to \Re$ is the *payoff function* of player $i$. We assume that the payoffs of the players are represented by some common monetary unit, and that the payoff functions are bounded[7].

A pre-game $G$ and a payoff function vector $U$ defines a game in strategic form denoted by $G(U)$. The game $G(U)$ is *finite* if the strategy sets are finite.

Let $x_i, y_i$ be strategies of player $i$ in the game $G(U)$.

$x_i$ *dominates* $y_i$ if $U_i(x_i, x_{-i}) \geq U_i(y_i, x_{-i})$ for every $x_{-i} \in X_{-i}$, and there exists $x_{-i} \in X_{-i}$ for which a strict inequality holds. $y_i$ is a *dominated* strategy if it is dominated by some other strategy of $i$. $x_i$ is a *dominant* strategy for $i$ if it dominates every other strategy of $i$. A profile of strategies $x$ is a (Nash) equilibrium if for every player $i$, $x_i$ is a best-response

---

5. Complexity of implementation when the organizer controls the structure of the game is discussed by Conitzer and Sandholm (2002).

6. The VCG mechanisms (Vickrey, 1961; Clarke, 1971; Groves, 1973) have been widely discussed in the context of combinatorial auctions, a topic which received much attention in the recent multi-agent systems and e-commerce literature, e.g., (Nisan, 2000; Sandholm, Suri, Gilpin, & Levine, 2001; Parkes, 1999)

7. If the game is finite the payoff functions are automatically bounded.

to $x_{-i}$. That is,

$$U_i(x_i, x_{-i}) \geq U_i(y_i, x_{-i}) \quad \text{for every } i \in N \text{ and } y_i \in X_i.$$

That is, if every player $i$ believes that all other players act according to $x$, he is better off by playing according to $x$. Modern economic theory has made an (some times implicit) assumption that economic interactions are in equilibrium. However, the rationale for this assumption is in debate in many cases, and it is particulary so when there exist multiple equilibrium profiles. In contrast, using a non-dominated strategy is a rational behavior in any reasonable definition of rationality. Moreover, refraining from the use of dominated strategies is taken as the most basic idea and agreed upon technique in decision theory.

Let $G = (N, X)$ be a pre-game. For every vector of payoff functions $V$, let $\bar{X}_i(V)$ be the set of non-dominated strategies of $i$ in the game $G(V)$, and let $\bar{X}(V) = \bar{X}_1(V) \times \bar{X}_2(V) \times \cdots, \bar{X}_n(V)$. $\bar{G}(V)$ is the game $(N, \bar{X}, V)$, where, by an innocent abuse of notations $V$ denotes the vector of the payoff functions restricted to $\bar{X}$. A vector $V$ of payoff functions is *non-negative* ($V \geq 0$) if $V_i(x) \geq 0$ for every player $i$ and for every $x \in X$.

Consider a set of *desired strategy profiles* $O \subseteq X$ in the game $G(U)$. A non-negative vector of payoff functions $V$ *implements* $O$ *in* $G(U)$ if

- $\emptyset \subset \bar{X}(U + V) \subseteq O$.

Such a $V$ is called a *k-implementation* of $O$ *in* $G(U)$, if in addition

- $\sum_{i=1}^{n} V_i(x) \leq k$ for every $x \in \bar{X}(U + V)$.

Obviously, by paying every player $i$ sufficient amount of money for playing the strategy associated with a particular strategy profile in $O$, one can implement $O$.

That is, the interested party commits herself to certain non-negative payoffs $V$, in such a way that "rational" players will only choose strategy profiles in $O$, and such that in the worst case the interested party will have to pay at most $k$.

Note that implicitly we have made two important assumptions :

- *Output observability*: The interested party can observe the actions chosen by the players.

- *Commitment power*: The interested party is reliable in the sense that the players believe that she will indeed pay the additional payoff defined by $V$.

However, the requirement $V \geq 0$ means that the interested party cannot force players to make payments based on their actions. In addition, the interested party cannot modify the set of available strategies, or enforce behavior in any way. He can just reliably promise positive monetary transfers conditioned on the observed outcome.

Let $k(O)$ be the price of implementing $O$. That is, $k(O)$ is the greatest lower bound (GLB) of all non-negative numbers $q$ for which there exists a $q$- implementation. That is, $k(O) = k$ implies that for every $\epsilon > 0$ $O$ has a $(k + \epsilon)$- implementation vector $V^\epsilon$, and $O$ does not have a $k'$- implementation for any $k' < k$. $V$ is an *optimal* implementation of $O$ if $V$ implements $O$ and

$$\max_{x \in \hat{X}(U+V)} \sum_{i=1}^n V_i(x) = k(O).$$

$V$ is an $\epsilon$ *optimal* implementation of $O$ if $V$ implements $O$ and

$$\max_{x \in \hat{X}(U+V)} \sum_{i=1}^n V_i(x) \le k(O) + \epsilon.$$

## 3. k-Implementation of singletons

When $O$ is a singleton, that is $O = \{z\}$, we sometimes abuse notations and we will say that $z$ (instead of $\{z\}$) has a $k$-implementation in $G(U)$, and we refer to $k(z)$ as the price of implementing $z$.

### 3.1 Finite games

In this section we focus on finite games, and on the characterization of optimal $k$ implementation of singletons.

**Theorem 1** *Let $G(U)$ be a finite game with at least two strategies to every player. Every strategy profile $z$ has an optimal implementation $V$, and moreover:*

$$k(z) = \sum_{i=1}^n \max_{x_i \in X_i} \left( U_i(x_i, z_{-i}) - U_i(z_i, z_{-i}) \right). \qquad 3.1.1$$

**Proof:** Let $z \in X$ and let $V$ implements $z$. Let $i \in N$. If for some $x_i \ne z_i$, for some $x_{-i}$, $V_i(x_i, x_{-i}) > 0$, then one can modify $V_i$ by changing this term to 0, and get a cheaper implementation of $z$. Hence, we can assume without loss of generality that we deal with payoff function vectors $V$ for which, for every $i$, $V_i(x_i, *) = 0$ for every $x_i \ne z_i$.

As $z_i$ is a dominant strategy for $i$ in $G(U + V)$,

$V_i(z_i, x_{-i}) + U_i(z_i, x_{-i}) \ge V_i(x_i, x_{-i}) + U_i(x_i, x_{-i})$ for every $x_i \in X_i$, and for every $x_{-i} \in X_{-i}$.

Since for $x_i \ne z_i$, $V_i(x_i, *) = 0$, a necessary condition for an implementation is

$$V_i(z_i, x_{-i}) + U_i(z_i, x_{-i}) \ge U_i(x_i, x_{-i}) \quad \text{for every } x_i \in X_i.$$

That is,

$$V_i(z_i, x_{-i}) \ge \max_{x_i \in X_i} \left( U_i(x_i, x_{-i}) - U_i(z_i, x_{-i}) \right).$$

One can use $x_{-i} \neq z_{-i}$ in order to get a costless strict inequality required by the definition of domination ( here we use our assumption that every player has at least two strategies). Hence, an optimal implementation vector for $z$, $V$ is defined for every $i$ by: $V_i(x_i, *) = 0$ for $x_i \neq z_i$, and $V_i(z_i, x_{-i}) = max_{x_i \in X_i}(U_i(x_i, z_{-i}) - U_i(z_i, z_{-i})) + \delta(x_{-i})$, where $\delta : X_{-i} \to \Re$ is a nonnegative function that satisfies $\delta(z_{-i}) = 0$, and for some $x_{-i} \neq z_{-i}$, $\delta(x_{-i}) > 0$. Therefore (3.1.1) is satisfied. $\square$

Note that $z$ is in equilibrium if and only if for every player $i$, $max_{x_i \in X_i}(U_i(x_i, z_{-i}) - U_i(z_i, z_{-i})) = 0$. Hence the following characterization of equilibrium is a corollary of Theorem 1:

**Corollary 1** *Let $G(U)$ be a finite game with at least two strategies to every player, and let $z \in X$. $z$ is in equilibrium if and only if $z$ has a zero- implementation.*

### 3.2 Infinite games

When the game $G(U)$ is infinite, one can get phenomena that contradicts our intuition. For example, it is possible that $\bar{X}_i = \{z_i\}$ but $z_i$ is not a dominant strategy. E.g., consider the two-person game in which player 1 can choose the strategy $z_1$, or any number $0 < x_1 < 1$, and player 2 can choose $z_2$ or $x_2$. $U_1(z_1, z_2) = 0.5, U_1(z_1, x_2) = 10, U_1(x_1, *) = x_1$. $U_2$ is an arbitrary function. It is easily seen that every $x_1$ is dominated by a bigger number in the open interval (0,1), $z_1$ is not dominated, and hence $\bar{X}_1 = \{z_1\}$. However, $z_1$ does not dominate $x_1$ for $x_1 > 0.5$. Moreover, the *max* operator used in the proofs of Theorem 1 and 2 may not be well-defined for infinite games. A game $G(U) = (N, X, U)$ is called *regular* if every $X_i$ is a compact metric space, and the payoff functions are continuous on $X$ endowed with the product metric.

**Theorem 2** *Theorems 1 holds for regular games.*

**Proof:** The proof requires very standard techniques, and therefore it is omitted. $\square$

We then immediately get:

**Corollary 2** *Corollary 1 holds for regular games.*

### 3.3 Mixed strategies

For every finite set $B$ we denote by $\Delta(B)$ the set of probability distributions over $B$. That is, $\Delta(B)$ consists of all functions $q : B \to [0, 1]$ with $\sum_{b \in B} q(b) = 1$. Let $G(U) = (N, X, U)$ be a finite game. The *mixed extension* of $G(U)$ is the infinite game $G^m(U^m) = (N, X^m, U^m)$, where $X^m = \Delta(X_1) \times \Delta(X_2) \times \cdots \times \Delta(X_n)$, and for every player $i$, $U_i^m(p_1, p_2, \cdots, p_n) = \sum_{x \in X} p_1(x_1) p_2(x_2) \cdots p_n(x_n) U_i(x)$. That is, $U_i^m(p)$ is the expected payoff of player $i$ when every player $j$ (including $i$) is choosing his strategy ( independently of the other players) with a randomizing device that chooses each strategy $x_j$ with probability $p_j(x_j)$.

A profile of mixed strategies $p \in X^m$ is called a *mixed-strategy equilibrium in $G(U)$* if $p$ is in equilibrium in the game $G^m(U^m)$. By Nash (1950) every finite game possesses a mixed strategy equilibrium. Note that every strategy $x_i \in X_i$ of $i$ can be identified with the mixed strategy in which $i$ chooses $x_i$ with probability 1. In this sense, $X_i$ is a subset of $X_i^m$. When we deal with an environment in which mixed strategies are considered, we refer to every strategy $x_i \in X_i$ as a *pure* strategy of $i$.

Note that the possibility of using mixed strategies does not destroy our previous results. That is, if $x_i$ is a dominant (dominated) strategy in $G(U)$, it continues to be a dominant (dominated) strategy in $G^m(U^m)$.

As $G^m(U^m)$ is a regular game we can apply Theorem 2 Corollary 2 and deduce:

**Theorem 3** *Let $G(U)$ be a finite game in strategic form with at least two strategies for every player. Let $p$ be a profile of mixed strategies in $G(U)$. $p$ is a mixed strategy equilibrium in $G(U)$ if and only if $p$ has a 0-implementation in $G^m(U^m)$.*

Hence, technically, the case of mixed strategies follows from the theorems regarding pure strategies in infinite games. However, the reader should notice that in this case our output observability assumption has a strong implication. Implementing a mixed strategy profile in $G^m(U^m)$ actually means *algorithm observability* in $G(U)$. That is, the interested party can observe the mixed strategies used by the players. This is a realistic assumption if we think about the interested party as a system's administrator that deploys algorithms submitted by users. The designer is not allowed to alter the users' algorithms, but can verify the exact content of these algorithms. Hence, for example, in such a setting, if a user's algorithm flips a coin in order to decide on its course of action, then the exact randomized algorithms, including the particular coin flipping, can be viewed by the interested party. The interesting case in which the interested party cannot observe the mixed strategies will be discussed in Section 6.

## 4. $k$-implementation of sets

In the previous sections we dealt with some properties of $k$- implementation. In particular we emphasized the interesting cases of k-implementations of singletons. However, from a computational perspective, given a game $G(U)$, and a set of desired strategy profile $O$, it may be of interest to find the smallest integer $k \geq 0$ for which a $k$- implementation exists. We can show:

**Theorem 4** *Given a game $G(U)$, a set of desired strategy profiles $O$, and an integer $k \geq 0$, deciding whether there exists a $k$ implementation of $O$ in $G(U)$ is NP-hard.*

**Proof:** In order to prove the above theorem, we use a reduction from the SAT problem. Given a set of primitive propositions $\{x_1, x_2, \ldots, x_n\}$, consider a CNF formula. A CNF formula is a conjunction of clauses $C_1 \wedge C_2 \wedge \ldots \wedge C_m$, where $C_i = l_{1_i} \vee l_{i_2} \vee \ldots \vee l_{i_{s_i}} (s_i \geq 2)$ and where $l_{j_q} = x_i$ or $l_{j_q} = \neg x_i$ for some $i$ (for every $1 \leq j \leq m$ and $1 \leq q \leq s_j$). The SAT problem is the following decision problem: given a CNF formula, is there a truth assignment to the primitive propositions that satisfies it? This problem is known to be NP-complete.

We will now show a polynomial reduction from SAT to the problem of deciding whether a 2- implementation exists, where the games are 2-person games. This will suffice to prove our result. Without loss of generality we restrict our attention to CNF formulas where both $x_i$ and $\neg x_i$ appear in the formula, and no clause refers to both $x_i$ and $\neg x_i$ ($1 \leq i \leq n$).

Both agents will have a strategy $c_i$ associated with clause $C_i$, for every $1 \leq i \leq m$. In addition, both agents will have strategies $y_i, z_i$, associated with the literals $x_i$ and $\neg x_i$, respectively ($1 \leq i \leq n$).

The payoff of agent 1, $p_1$, will be defined as follows. For any strategy profile of the form $(c_i, y_j)$ the payoff will be 3 if $x_j$ appears in clause $i$ and 0 otherwise. For any strategy profile of the form $(c_i, z_j)$ the payoff will be 3 if $\neg x_j$ appears in clause $i$ and 0 otherwise. For any strategy profile of the form $(c_i, c_j)$ the payoff will be 50 if $i = j$ and 0 otherwise. For any strategy profile of the form $(y_i, y_j)$ the payoff will be 2 if $i = j$ and 3 otherwise. For any strategy profile of the form $(z_i, z_j)$ the payoff will be 2 if $i = j$ and 3 otherwise. For any strategy profile of the form $(y_i, z_j)$ or the form $(z_i, y_j)$ the payoff will be 1 if $i = j$ and 3 otherwise. For any strategy profile of the form $(y_i, c_j)$ the payoff will be 51 if $x_i$ or $\neg x_i$ appear in clause $C_j$ and 0 otherwise. For any strategy profile of the form $(z_i, c_j)$ the payoff will be 51 if $x_i$ or $\neg x_i$ appears in clause $C_j$ and 0 otherwise.

The payoff to agent 2, $p_2$, will be as follows. The payoff for any strategy profile of the form $(y_i, y_i)$ or $(z_i, z_i)$ will be 101; the payoff for any strategy profile of the form $(y_i, z_i)$ or $(z_i, y_i)$ will be 100; the payoff for any strategy profile of the form $(y_i, z_j)$ or $(z_i, y_j)$ where $i \neq j$ will be 0. The payoff for any strategy profile of the form $(c_i, c_j)$ will be 50 if $i = j$, and 0 otherwise. For any strategy profile of the form $(y_i, c_j), (z_i, c_j)$ the payoff will be 50 if $i = j$ and 0 otherwise. The payoff for any strategy profile of the form $(c_i, y_j), (c_i, z_j)$ will be 0.

The set $O$ of desired strategy profiles will include all strategy profiles excluding the following: All strategy profiles of the form $(c_i, s), (y_i, z_i), (z_i, y_i)$ (where $s$ is any strategy) will be prohibited.

If the formula is satisfiable then there is 2- implementation: add 1 to the payoff obtained by agent 1 in any strategy profile of the form $(y_i, s)$ if $x_i$ is true in the satisfying assignment, and add 1 to the payoff obtained by agent 1 in any strategy profile of the form $(z_i, s)$ if $x_i$ it false. As for agent 2, increase its payoff by 1 for $(z_i, y_i)$ if $x_i$ is true, and increase its payoff by 1 for $(y_i, z_i)$ if $x_i$ is false.

Notice that given the above construction the strategies of the form $c_i$ of agent 1 will become dominated and will be removed. In addition, if $x_i$ has been assigned true (resp. false) then strategy $z_i$ (resp. $y_i$) will become dominated. The corresponding strategies of agent 2 (i.e. $z_i$ if $x_i$ is true and $y_i$ if $x_i$ is false) will become dominated too, which yield the desired behavior.

Similarly, notice that since we must remain with at least one $y_i$ or $z_i$ for agent 1, it must be the case that at least one $y_i$ or $z_i$ will be removed for agent 2, and that this cannot be obtained with a payment of 1 (increasing the payoff from 100 to 101). Hence, in a 2- implementation the payment of agent 2 should be 1. However, notice that we must add 1 to the payoff of agent 1 for at least one of the elements of the form $(y_i, y_i)$ or $(z_i, z_i)$ which correspond to an $x_i$ or $\neg x_i$ that appear in clause $j$ (i.e. for at least one of the literals in clause $j$, we need to add 1 to a strategy profile of the form $(y_i, y_i)$ or $(z_i, z_i)$ associated with that literal). Notice that this is based on the fact that the strategy $c_i$ of agent $i$ cannot be removed by adding a payment of 1 to the outcome of any strategy of the form $y_j, z_j, c_j$ of agent 1, where $x_j$ and $\neg x_j$ do not appear in $C_i$, since the payoff of agent 1 for $(y_j, c_i)$ and for $(z_j, c_i)$ is 0, while the payoff of agent 1 for $(c_i, c_i)$ is 50. Moreover, if we add 1 to the payoff that agent 1 obtains for both $(y_i, y_i)$ and $(z_i, z_i)$ then both $y_i$ and $z_i$ will not be dominated for agent 1, which will result in the possibility of playing a strategy profile which is not desired(given that it is impossible to remove both $y_i$ and $z_i$ for agent 2). Hence, the implementation corresponds to a sound truth assignment to the CNF formula, where $x_i$ is assigned true iff the payoff for agent 1 at $(y_i, y_i)$ has been augmented by 1.

∎

Notice that the previous result applies already in the case where we have a constant number of agents. The previous result suggests one may wish to consider relaxations of the optimal implementation problem that will be tractable.[8] One interesting relaxation[9] is the following one:

A non-negative vector of payoff functions $V$ is called a *k- exact implementation* of $O$ in $G(U)$, if the following two conditions are satisfied:

- $\bar{X}(U + V) = O$.

- $\sum_{i=1}^{n} V_i(x) \leq k$ for every $x \in O$.

Hence, $V$ implements $O$ means that the set of non-dominated strategies in $G(U + V)$ is a subset of $O$, while $V$ is an exact implementation of $O$ if this set equals $O$. When dealing with singletons the concepts of implementation and exact implementation coincide.

Notice that the concept of exact implementation makes sense only when $O = O_1 \times O_2 \cdots \times O_n \subseteq X = X_1 \times X_2 \cdots \times X_n$ since otherwise it will be impossible to (exactly) implement $O$. We also assume that $O_i$ is strictly contained in $X_i$ for every agent $i$, and that $O_i$ does not contain two strategies such that one dominates the other. We can show:

**Theorem 5** *Computing the optimal $k$ for which an exact implementation exists is polynomial.*

The algorithm leading to the above result is now illustrated for the case of two agents. We construct the game matrix $G'$, where the payoff function of agent $i$ is denoted by $p_i$; $p_i$ describes the payment to agent $i$ for the different strategy profiles (if/when selected). The matrix $G'$ will be the matrix of perturbations (non-negative monetary promises), while $\bar{G}$ will denote the perturbed matrix generated. Let $M = K + 1$ where $K$ is the maximal element in the original game matrix.

**The optimal perturbation [OP] algorithm:**

---

8. Another approach may be to search for good approximation techniques.
9. See the discussion in the last section.

1. Let $(e_1, \ldots, e_k)$ the list of possible differences between an agent's payoffs in the original game (i.e. the possible results one obtains by subtracting two possible payoffs of an agent in the given game) , sorted from small to large.

2. Let $p_1(a, b) := M$ for every $a \in O_1$ and $b \in X_2 \setminus O_2$, and let $p_1(a, b) = 0$ whenever $a \in X_1 \setminus O_1$ or $b \in O_2$.

3. Let $p_2(a, b) := M$ for every $b \in O_2$ and $a \in X_1 \setminus O_1$, and let $p_2(a, b) := 0$ whenever $b \in X_2 \setminus O_2$ or $a \in O_1$,

4. Let i:=1

5. Let $e := e_i$

6. Let $p_1(a, b) := e$ for every strategy profile of the form $(a, b)$ where $a \in O_1$ and $b \in O_2$

7. Let $\bar{G} := G + G'$

8. If the non-dominated strategies for agent 1 in $\bar{G}$ do not coincide with $O_1$ then let i:=i+1 and return to 5

9. Let i:=1

10. Let $e := e_i$

11. Let $p_2(a, b) := e$ for every $a \in O_1$ and $b \in O_2$

12. Let $\bar{G} = G + G'$

13. If the non-dominated strategies for agent 2 in $\bar{G}$ do not coincide with $O_2$ then let i:=i+1 and return to 10

## 5. Incomplete information

In previous sections we dealt with games with complete information. However, in many real life situations the players ( and the interested party) have incomplete information about certain parameters of the game. In the economic literature this phenomenon has been mainly modelled by a Bayesian setting. In this setting every player receives some private signal, which is correlated with the unknown parameters, and the joint distribution of signals is commonly known to all players (and to the interested party). In the following subsection we deal with Bayesian games without probabilistic information. Such games are called *games in informational form.*

### 5.1 Games in informational form

The precise definition of games in informational form will not be given in this paper, in which we focus on a particular type of such games – combinatorial auctions. However, a typical example is shown in Figure 1.

**Figure 1: A game in informational form**

In this game, Player 1 is about to receive one of the signals $s_1$ or $t_1$, and Player 2 is about to receive one of the signals[10], $s_2$ or $t_2$. The true game to be played is determined by the pair $(c_1, d_2)$, where $c, d \in \{s, t\}$. However, neither player knows the exact game. Given $s_1$ $(s_2)$ player 1 (player 2) has to choose an action in $\{U_1, D_1\}$ $(\{L_1, R_1\})$, and given $t_1$ $(t_2)$ player 1 (player 2) has to choose an action in $\{U_2, D_2\}$ $(\{L_2, R_2\})$. The payoffs are shown in the figure. A Bayesian game is obtained from a game in informational form by adding a probability distribution over the pairs of signals as described in Figure 2. The probability that 1 receives the signal $c_1$, and 2 receives $d_2$ equals $p_{ij}$, where $i = 1$ if $c = s$, $i = 2$ if $c = t$, $j = 1$ if $d = s$, $j = 2$ if $d = t$.

**Figure 2: A Bayesian Game**

A strategy of a player is a function defined on the set of signals, which assigns to every signal an action[11] in the games that are consistent with this signal. For example, in the game in Figure 1, a strategy of player 1 is a function $b_1 : \{s_1, t_1\} \to \{U_1, D_1, U_2, D_2\}$, with the property that $b_1(s_1) \in \{U_1, D_1\}$ and $b_1(t_1) \in \{U_2, D_2\}$. A strategy of player 2 is analogously defined as a function $b_2 : \{s_2, t_2\} \to \{L_1, R_1, L_2, R_2\}$. The concepts of domination and of equilibrium ( traditionally referred to as ex post equilibrium) are naturally defined. For example, in Figure 3 The strategy of player 1 in which she chooses $U_1$ when she receives the signal $s_1$, and she chooses $D_2$ given $t_1$ dominates each of the other four strategies of player 1.

Figure 3: Domination

That is, given $s_1$, independently of the other player's signal and action, choosing $U_1$ is at least as good as choosing $D_1$, and for at least one signal and action of Player 2, choosing $U_1$ is strictly better than choosing $D_1$.

Figure 4 demonstrate an ex post equilibrium.

---

11. In an environment in which complex strategies exist, we refer to the choices of a player at a game in strategic form as actions.

**Figure 4: Ex Post Equilibrium**

Note that even under the output observability assumption a player does not have to reveal her strategy. The signal of a player is her private knowledge, and she reveals only the action she chooses. The interested party does not observe the signals; therefore, if the action sets in all games are the same ( as is the case in Figure 4) the interested party does not receive any information about the true game to be played. Hence, the only thing he can do is to use the same vector $V$ at all games. Therefore:

**Claim** For every $k \geq 0$ It is impossible to $k$-implement the ex post equilibrium described in Figure 4.

**Proof** The interested party wishes to make $U$ a dominant strategy at the game $(s_1, t_2)$, and she wishes to make $D$ a dominant strategy at $(t_1, t_2)$. Assume $V_1(U, L) = x$ and $V_1(D, L) = y$, then the following two contradictory inequalities should be satisfied: $0 + x \geq 7 + y$, and $5 + x \leq 4 + y$. □

As is stated in the next subsection, when the game in informational form has a particular structure, our results for the complete information case are generalized.

## 5.2 The VCG combinatorial auctions

Combinatorial auctions constitute a special class of games in informational form. Our notations and definitions are taken from Holzman and Monderer (2002).

In a combinatorial auction there is a seller, denoted by 0, who wishes to sell a set of $m$ goods $A = \{a_1, \ldots, a_m\}$, $m \geq 1$, that she owns. We denote by $2^A$ the family of all bundles of goods (i.e., subsets of $A$). There is a set of $n$ buyers $N = \{1, \ldots, n\}$, $n \geq 1$. An *allocation* of the goods is an ordered partition $\gamma = (\gamma_0, \gamma_1, \ldots, \gamma_n)$ of $A$.[12] We denote by $\Gamma$ the set of all allocations.

---

12. Note that the goods are allocated among the buyers and the seller. We assume, however, that the seller derives no utility from keeping any of the goods, and that she does not set strategic reserve prices.

A buyer's *valuation function* is a function $v : 2^A \rightarrow \Re$, satisfying $v(\emptyset) = 0$ and

$$B \subseteq C, \ B, C \in 2^A \Rightarrow v(B) \leq v(C).$$

When buyer $i$ with the valuation function $v_i$ receives the set of goods $B$, and pays a monetary transfer $c_i \in \Re$ his utility is $v_i(B) - c_i$. Every buyer knows his valuation function.

We denote by $V$ the set of all possible valuation functions. The set $V^N$, the $n$-fold product of the set $V$, is the set of all profiles of valuations $\mathbf{v} = (\mathbf{v_1}, \ldots, \mathbf{v_n})$, one for each buyer.

For an allocation $\gamma = (\gamma_0, \gamma_1, \ldots, \gamma_n) \in \Gamma$ and a profile of valuations $\mathbf{v} = (\mathbf{v_1}, \ldots, \mathbf{v_n}) \in \mathbf{V^N}$ we denote by $S(\mathbf{v}, \gamma)$ the *total social surplus* of the buyers, that is,

$$S(\mathbf{v}, \gamma) = \sum_{\mathbf{i} \in \mathbf{N}} \mathbf{v_i}(\gamma_{\mathbf{i}}).$$

We also denote

$$S_{max}(\mathbf{v}) = \max_{\gamma \in \mathbf{\Gamma}} \mathbf{S}(\mathbf{v}, \gamma),$$

and we refer to an allocation $\gamma$ that achieves this maximum as an *optimal* allocation for $\mathbf{v}$.

A *Vickrey-Clarke (VC)* auction mechanism is described as follows. Every buyer $i$ is required to report a valuation function $\widehat{v}_i$. Based on the reported valuations $\widehat{\mathbf{v}} = (\widehat{v}_1, \ldots, \widehat{v}_n) \in V^N$ the mechanism selects an allocation $d(\widehat{\mathbf{v}}) = (d_0(\widehat{\mathbf{v}}), \ldots, d_n(\widehat{\mathbf{v}})) \in \Gamma$, which is optimal for $\widehat{\mathbf{v}}$. Because ties are possible, such an allocation may not be unique, and therefore there is more than one VC mechanism. Every function $d : V^N \rightarrow \Gamma$ satisfying $S(\widehat{\mathbf{v}}, \mathbf{d}(\widehat{\mathbf{v}})) = \mathbf{S_{max}}(\widehat{\mathbf{v}})$ for all $\widehat{\mathbf{v}} \in \mathbf{V^N}$ determines uniquely a VC mechanism, which we refer to as the VC mechanism $d$. This mechanism assigns to buyer $i$ the bundle $d_i(\widehat{\mathbf{v}})$ and makes him pay $c_i^d(\widehat{\mathbf{v}})$ to the seller, where

$$c_i^d(\widehat{\mathbf{v}}) = \max_{\gamma \in \Gamma} \sum_{j \neq i} \widehat{v}_j(\gamma_j) - \sum_{j \neq i} \widehat{v}_j(d_j(\widehat{\mathbf{v}})).$$

This represents the loss to the other agents' total surplus caused by agent $i$'s presence.

A *Vickrey-Clarke-Groves (VCG)* auction mechanism is parameterized by a VC mechanism $d$, and by an $n$-tuple $\mathbf{h} = (\mathbf{h_1}, \ldots, \mathbf{h_n})$ of functions $h_i : V^{N \setminus \{i\}} \rightarrow \Re$. The mechanism selects an allocation according to the allocation function $d$, and the transfer function of buyer $i$ is

$$c_i^{d, \mathbf{h}}(\widehat{\mathbf{v}}) = c_i^d(\widehat{\mathbf{v}}) + h_i(\widehat{\mathbf{v}}_{-i}).$$

Hence, a VC auction mechanism is a special type of VCG auction mechanism, in which $h_i$ is the function that is identically equal to zero for every $i$.

Let $AM = (d, h)$ be a VCG mechanism. The utility of $i$ with the valuation $v_i$ depends on the vector of reported valuations $\widehat{\mathbf{v}} = (v_i, \widehat{\mathbf{v}}_{-i})$, and it is denoted by $u_i(v_i, \widehat{v}_i, \widehat{\mathbf{v}}_{-i})$. That is,

$$u_i(v_i, \widehat{v}_i, \widehat{\mathbf{v}}_{-i}) = v_i(d_i(\widehat{\mathbf{v}})) - c_i^{d, h}(\widehat{\mathbf{v}}).$$

The behavior of buyer $i$ in a mechanism $AM$ is described by a *strategy* $b_i : V \rightarrow V$.

A strategy $b_i$ is a *dominant* strategy for $i$ if the following two conditions hold:[13]

- For every $v_i \in V$, and for every $\widehat{\mathbf{v}}_{-i} \in V^{N \setminus \{i\}}$

$$u_i(v_i, b_i(v_i), \widehat{\mathbf{v}}_{-i}) \geq u_i(v_i, \widehat{v}_i, \widehat{\mathbf{v}}_{-i}) \quad \text{for every } \widehat{v}_i \in V.$$

- For every $v_i \in V$, there exists $\widehat{\mathbf{v}}_{-i} \in V^{N \setminus \{i\}}$ such that

$$u_i(v_i, b_i(v_i), \widehat{\mathbf{v}}_{-i}) > u_i(v_i, \widehat{v}_i, \widehat{\mathbf{v}}_{-i}).$$

A strategy profile $(b_1, \ldots, b_n)$ forms an *ex post equilibrium* if for every profile of valuations $\mathbf{v} = (\mathbf{v_1}, \ldots, \mathbf{v_n}) \in \mathbf{V^N}$, and for every buyer $i$,

$$u_i(v_i, b_i(v_i), \mathbf{b_{-i}(v_{-i})}) \geq \mathbf{u_i(v_i, \widehat{v}_i, b_{-i}(v_{-i}))} \quad \text{for every } \widehat{v}_i \in V,$$

where $\mathbf{b_{-i}(v_{-i})} = \mathbf{(b_j(v_j))_{j \neq i}}$. The profile $(b_1, \ldots, b_n)$ is *symmetric* if $b_i = b_j$ for every two buyers $i, j \in N$.

It is well-known that every VCG auction mechanism is truth-telling in the sense that for every buyer $i$, the strategy $b_i(v_i) = v_i$ of revealing the true valuation is a dominant strategy.[14]

Special type of strategies were considered by Holzman et al. (2003), Holzman and Monderer (2002). A *bundling strategy* for buyer $i$ is parameterized by a subfamily $\Sigma_i$ of $2^A$ such that $\emptyset \in \Sigma_i$, and is denoted by $f^{\Sigma_i}$. It maps every $v \in V$ to $v^{\Sigma_i} \in V$ defined by

$$v^{\Sigma_i}(B) = \max_{C \subseteq B, C \in \Sigma_i} v(C) \quad \text{for every } B \in 2^A.$$

This has the effect of pretending that the agent cares only about bundles in $\Sigma_i$ (for which he announces his true valuation), and derives his valuation for other bundles by maximizing over the bundles in $\Sigma_i$ that they contain.

A valuation $v^{\Sigma_i}$ that satisfies the above equalities is said to be *based on* $\Sigma_i$ ( or, simply $\Sigma_i$-based). The set of all $\Sigma$-based valuation function is denoted by $V^{\Sigma}$.

A subfamily $\Sigma$ of $2^A$ such that $\emptyset \in \Sigma$ is a *quasi field* if it satisfies the following two conditions:

$$B \in \Sigma \Rightarrow A \setminus B \in \Sigma,$$

$$B, C \in \Sigma \text{ and } B \cap C = \emptyset \Rightarrow B \cup C \in \Sigma.$$

In the work by Holzman and Monderer (2002) it was proven that every ex post equilibrium in the VCG mechanisms is a bundling equilibrium in the following sense: For every

---

$n \geq 3$, every profile $(b_1, b_2, \cdots, b_n)$ of strategies, which satisfies that any subprofile $(b_i)_{i \in N'}$, $N' \subseteq N$, is an ex post equilibrium in all VCG mechanisms is a symmetric profile of bundling strategies. That is, there exists $\emptyset \in \Sigma \subseteq 2^A$ such that $b_i = f^\Sigma$ for every $i \in N$, and moreover, it was proved by Holzman et al. (2003) that this $\Sigma$ must be a quasi field.

### 5.3 0-Implementation of ex post equilibrium in frugal VCG auction mechanisms

We begin with a formal definition of a frugal VCG combinatorial auction:

**Definition of Frugal VCG mechanisms** A VCG mechanism $(d, h)$ is called *frugal* if $d$ does not allocate unnecessary goods to the buyers. That is, for every $\widehat{v} = (\widehat{v}_1, \ldots, \widehat{v}_n) \in V^N$ the mechanism selects an allocation $d(\widehat{v}) = (d_0(\widehat{v}), \ldots, d_n(\widehat{v})) \in \Gamma$, which is optimal for $\widehat{v}$ and satisfies in addition:

- For every player $i$, and for every $B_i \subset \gamma_i$,

$$\widehat{v}_i(B_i) < \widehat{v}_i(\gamma_i),$$

  where $\gamma_i = d_i(\widehat{v})$.

Intuitively, in a frugal VCG mechanism the center never allocates unnecessary goods. If an agent has a bid for a bundle $B_1$ and the same bid for a superset of it, $B_2$, the center will never allocate $B_2$ to that agent (the bid for $B_2$ should be strictly higher than the bid for $B_1$ in order that allocating $B_2$ to that agent will be a possibility.)

Consider an interested party who wishes to 0-implement the ex post equilibrium $b = (b_1, b_2, \cdots, b_n)$ in the VCG auction mechanism $(d, h)$. Given the result by Holzman and Monderer (2002) stated at the end of the previous subsection we can assume almost without loss of generality that $b_i = f^\Sigma$ for every $i \in N$. The interested party wishes to promise a positive payment to every buyer $i$ whenever he follows the recommendation to play according to $b_i$, and at least one of the players, say $j$, does not play according to $b_j$. However, the interested party does not know the valuation functions. Hence, how could she know whether a player follows the recommendation? Indeed she cannot. However, since $\Sigma$ is known to the interested party she can partially monitor the players strategies, because, independently of a player's valuation function his reported valuation function must be $\Sigma$-based. Hence, the best the interested party can do is to offer every player $i$ a positive payment if his reported valuation is $\Sigma$-based, and at least one of the other players' reported a valuation function, which is not $\Sigma$-based. These payments can be made arbitrarily high so that reporting a $\Sigma$-based valuation function will yield a higher payoff for $i$ than any other, non $\Sigma$-based valuation function if at least one of the other players does not report a $\Sigma$-based valuation function. However, player $i$ can cheat within the set of $\Sigma$-based valuations without being caught! It turns out that when the VCG mechanism is frugal, every player is better off not cheating.

**Lemma 1** *Let $(d, h)$ be a frugal VCG mechanism, let $\Sigma$ be a quasi field, and let $i \in N$. For every profile of reported valuations of the other players, $\widehat{v}_{-i}$*

$$u_i(v_i, v_i^\Sigma, \widehat{v}_{-i}) \geq u_i(v_i, w_i, \widehat{v}_{-i}) \quad \text{for every } w_i \in V^\Sigma.$$

**Proof:** Without loss of generality we can assume that $h_j$ is constantly 0 for every $j$. Hence, the VCG mechanism is actually a VC mechanism. If $\widehat{v}_j$ is $\Sigma$-based for every $j$, $j \neq i$, the inequality follows from the fact that $f^\Sigma$ induces an ex post equilibrium. However, our proof does not use this fact. Let $\gamma = d(v_i^\Sigma, \widehat{\mathbf{v}}_{-i})$ be the allocation chosen by the auctioneer when $i$ reports $v_i^\Sigma$, and let $\delta = d(w_i, \widehat{\mathbf{v}}_{-i})$. Denote

$$t = \max_{\gamma \in \Gamma} \sum_{j \neq i} \widehat{v}_j(\gamma_j).$$

As the VCG mechanism is frugal, we have that $\gamma_i \in \Sigma$ and $\delta_i \in \Sigma$. Therefore we have that $v_i^\Sigma(\gamma_i) = v_i(\gamma_i)$, and $v_i^\Sigma(\delta_i) = v_i(\delta_i)$. By definition we have that $u_i(v_i, v_i^\Sigma, \widehat{\mathbf{v}}_{-i}) = S(v_i, \widehat{\mathbf{v}}_{-i}, \gamma) - t$. Now since $v_i^\Sigma(\gamma_i) = v_i(\gamma_i)$ we have that $u_i(v_i, v_i^\Sigma, \widehat{\mathbf{v}}_{-i}) = S(v_i^\Sigma, \widehat{\mathbf{v}}_{-i}, \gamma) - t$. By the optimality of $\gamma$ we get that $S(v_i^\Sigma, \widehat{\mathbf{v}}_{-i}, \gamma) - t \geq S(v_i^\Sigma, \widehat{\mathbf{v}}_{-i}, \delta) - t$. However, since frugality also implies that $v_i^\Sigma(\delta_i) = v_i(\delta_i)$, we get the following equation and the desired inequality:

$$S(v_i^\Sigma, \widehat{\mathbf{v}}_{-i}, \delta) - t = S(v_i, \widehat{\mathbf{v}}_{-i}, \delta) - t = u_i(v_i, w_i, \widehat{\mathbf{v}}_{-i}). \square$$

.

Hence, in the proof of Lemma 1 we used the fact that a frugal VCG mechanism must allocate a subset of goods in $\Sigma$ to every player who reports a $\Sigma$-based valuation function. The next example shows that Lemma 1 does not hold for an arbitrary VCG mechanism.

**Example**

There are two buyers and four goods a,b,c,d.

$$\Sigma = \{\emptyset, ab, cd, abcd\}.$$

The valuation function of 1 is $v_1$, and the reported valuation of 2 is $\widehat{v}_2$. Consider a VC auction mechanism that allocate $ab$ to 1 and $cd$ to 2 when 1 reports $v_1^\Sigma$, and it allocates $abc$ to 1, and $d$ to 2, when 1 reports the $\Sigma$-based valuation function $w_1$. In both cases 1 pays 0. Hence, reporting $v_1^\Sigma$ yields a utility of 1, while cheating yields 1.1. Note that the VC mechanism is not frugal because when 1 declares $w_1$ he receives $abc$, and $w_1(ab) = w_1(abc)$.

|  | $\emptyset$ | $a$ | $b$ | $c$ | $d$ | $ab$ | $ac$ | $ad$ | $bc$ | $bd$ | $cd$ | $abc$ | $abd$ | $acd$ | $bcd$ | $abcd$ |
|---|---|---|---|---|---|---|---|---|---|---|---|---|---|---|---|---|
| $v_1$ | 0 | 0 | 0 | 0 | 0 | 1 | 0 | 0 | 0 | 0 | 1.1 | 1 | 0 | 0 | 1.1 |
| $\widehat{v}_2$ | 0 | 0 | 0 | 0 | 0.75 | 0 | 0 | 0.75 | 0 | 0.75 | 0.75 | 0 | 0.75 | 0.75 | 0.75 | 0.75 |
| $v_1^\Sigma$ | 0 | 0 | 0 | 0 | 0 | 1 | 0 | 0 | 0 | 0 | 0 | 1 | 1 | 0 | 0 | 1.1 |
| $w_1$ | 0 | 0 | 0 | 0 | 0 | 1 | 0 | 0 | 0 | 0 | 0.1 | 1 | 1 | 0.1 | 0.1 | 1.1 |

We need the following terminology. Let $(d, h)$ be a VCG combinatorial auction, and let $\bar{M} > 0$. We denote by $(d, h, \bar{M})$ the direct combinatorial auction with the rules induced by $(d, h)$ in which, the set of feasible valuation functions ( and the set of bids) is $V(\bar{M})$, which is the set of all valuation functions $v$ satisfying $v(A) < \bar{M}$. The assumption of an upper bound is natural but not common in the literature of mechanism design. It can be verified that Lemma 1 holds for a VCG combinatorial auctions with bounded valuation functions.

**Theorem 6** *Let* $(d, h, \bar{M})$ *be a frugal VCG auction mechanism with at least two buyers. Let* $\Sigma$ *be a quasi fields of bundles. Then the symmetric ex post equilibrium induced by* $\Sigma$ *is 0-implementable.*

**Proof:** For every player $i$, the interested party promises $i$ a very high payoff ( e.g., $2\bar{M}+1$ if he reports a valuation function in $V^{\Sigma}$, and at least one of the other players does not report a $\Sigma$-based valuation function. Player $i$ is promised no payment if all players report valuation functions in $V^{\Sigma}$. By Lemma 1, $f^{\Sigma}$ is a dominant strategy for every player.□

## 6. Implementation devices

As we mentioned in Section 3, the proof of our result (Theorem 3) that every mixed strategy equilibrium is 0-implementable relies on the assumption that the interested party observes the mixed strategies used by the players. In this section we prove this result without this assumption. That is, the interested party can observe only the actions generated by the mixed strategies, but not the algorithms that generate them. In order to deal with this issue we define a new type of implementation by an implementation device.

Let $G(U) = (N, X, U)$ be a finite game in strategic form. An *implementation device for* $G(U)$ is a tuple $I = (S, h, \tilde{V})$, where $S = S_1 \times S_2 \cdots \times S_n$, $h \in \Delta(S)$ is a probability distribution over $S$, and $\tilde{V} : S \times X \to \Re_+^n$. $S_i$ is the finite[15] set of signals that can be sent to $i$. The interested party uses the implementation device $I$ as follows: She makes the device public, she secretly runs a randomizing scheme that chooses every $s \in S$ with a probability $h(s)$. If $s = (s_1, s_2, \cdots, s_n)$ is chosen, she sends player $i$ the signal $s_i$. If the strategy profile $x$ is selected then agent $i$ is paid $\tilde{V}_i(s, x)$. The implementation device generates a new game $G(U, I)$. This is actually a Bayesian game. A strategy for $i$ in this game is a function $b_i : S_i \to X_i$. For every $s_i$ and $x_i$, and a vector $b_{-i}$ of the other players let $W_i(x_i|s_i, b_{-i})$ be the expected payoff of $i$ in the game $G(U, I)$ if it chooses $x_i$ given that it receives the signal $s_i$ and all other players use $b_{-i}$. That is,

$$W_i(x_i|s_i, b_{-i}) = E_{s_{-i}}\left(U_i(x_i, b_{-i}(s_{-i})) + V_i(s_i, s_{-i}, x_i, b_{-i}(s_{-i})|s_i)\right),$$

where $s_{-i} = (s_j)_{j \neq i}$, and $b_{-i}(s_{-i}) = (b_j(s_j))_{j \neq i}$. A strategy $b_i$ is a dominant strategy for $i$ if for every signal $s_i$ with a positive probability ( that is $\sum_{t \in S, t_i = s_i} h(t) > 0$), and for every $b_{-i}$

$$W_i(b_i(s_i)|s_i, b_{-i}) \geq W_i(x_i|s_i, b_{-i}) \quad \text{for every } x_i \in X_i,$$

and there exists a profile $b_{-i}$ of the other players for which a strict inequality holds.

Every profile $b = (b_i)_{i \in N}$ determines a probability distribution $prob_b$ over $X$ defined as follows:

$$prob_b(x) = h(b_1 = x_1, b_2 = x_2, \cdots, b_n = x_n).$$

---

15. If $S$ is an infinite set, we must specify additional parameters required in probability theory. We associate with each $S_i$ a $\sigma$-algebra of events $\Sigma_i$, we endow $S$ with the product $\sigma$-algebra, $\Sigma$, and we define $h$ over $\Sigma$.

Let $\xi$ be a *desired probability distribution* over $X$. We say that $I$ *implements* $\xi$ in $G(U)$, if in $G(U, I)$ every player $i$ has a dominant strategy $b_i$, and $prob_b = \xi$. We say that $I$ is a *k-implementation of $\xi$ in $G(U)$* if $I$ implements $\xi$, and for every $s$ with $h(s) > 0$,

$$\sum_{i=1}^{n} \tilde{V}_i(s_i, b_i(s_i)) \leq k.$$

### 6.1 Mixed Strategies: Removing the algorithm observability assumption

Let $p = (p_1, p_2, \cdots, p_n)$ be a mixed strategy profile in a finite game $G(U)$. $p$ generates a probability distribution $\xi_p$ over $S$ as follows:

$$\xi_p(x) = p_1(x_1)p_2(x_2)\cdots p_n(x_n) \quad \text{for every } x \in X.$$

We say that an implementation device $I$ implements $p$ if it implements $\xi_p$.

In order to implement a mixed strategy equilibrium , the interested party employs an implementation device $I$, in which the set of signals $S_i$ is the set of actions of $i$, $X_i$. $h$ is the product probability of $p$. That is, $h(x) = p_1(x_1)p_2(x_2)\cdots p_n(x_n)$, and the function $\tilde{V}_i : S_i \times X_i \to \Re_+$ is designated in such a way that the strategy $b_i(s_i) = s_i$, $s_i \in X_i$, is a dominant strategy for every player $i$.

Hence, the interested party flips a coin for each player $i$ according to the probability $p_i$ in the profile she wishes to implement, and she sends the outcome of this coin flipping to $i$. Thus, the signals sent to the players are just recommendations to play. The payoff functions $\tilde{V}_i$, $i \in N$ are designed in such a way that obeying the recommendation is a dominant strategy for every player.

**Theorem 7** *Let $G(U)$ be a finite game with at least two actions for every player. Every mixed strategy equilibrium profile $p$ is 0-implementable in $G(U)$ with an appropriate implementation device $I = (S, h, \tilde{V})$ in which $S = X$, and $h = \xi_p$ is the product probability on $X$ defined by $p$.*

**Proof:** Denote by $b_i$ the strategy of $i$ in which it obeys every recommendation. That is $b_i(s_i) = s_i$ for every $s_i \in X_i$. The function $\tilde{V}_i$ should satisfy for every vector $d_{-i}$ of the other players' strategies,

$$W_i(s_i | s_i, d_{-i}) \geq W_i(x_i | s_i, d_{-i}) \quad x_i \in X_i.$$

We can assume without loss of generality that for every player $i$, $\tilde{V}_i(s_i, s_{-i}, x_i, y_{-i}) = 0$ for every $x_i \neq s_i$. Therefore the above inequality can be written as follows:

$$E_{s_{-i}}\left(\tilde{V}_i(s_i, s_{-i}, s_i, d_{-i}(s_i))\right) \geq E_{s_{-i}}\left(\tilde{U}_i(x_i, d_{-i}(s_i)) - U_i(s_i, d_{-i}(s_i))\right).$$

If $d_{-i} = b_{-i}$, that is $d_j(s_j) = s_j$ for every $j \neq i$ and for every $s_j \in X_j$, the right-hand-side of the above inequalities are non-positive because $p$ is a mixed strategy equilibrium. Hence, we may define $\tilde{V}_i(s, s) = 0$ for every $s \in X$. To make sure that the inequalities hold in all other cases (i.e., for all $d_{-i}$), we can define $\tilde{V}_i(s, s_i, x_{-i}) = 2M + 1$ for every $x_{-i} \neq s_{-i}$, where $M > 0$ is an upper bound on the absolute value of all players' payoff functions. The choice of $2M + 1$ (rather then $2M$) ensures the existence of a strict inequality required by the definition of domination.$\square$

## 6.2 0-implementations of correlated equilibrium

Aumann (1974) introduced the concept of correlated equilibrium. We provide one of the many equivalent definitions:

**Definition** Let $G(U) = (N, X, U)$ be a finite game in strategic form. A *correlated equilibrium* of $G(U)$ is a probability distribution $\xi$ over $X$ ($\xi \in \Delta(X)$) such that the strategies $b_i(s_i) = s_i$, $s_i \in X_i$, $i \in N$, form an equilibrium in the game $G(U, I)$, where $I = (S, h, \tilde{V})$ is the following implementation device:

- $S = X$,

- $h = \xi$,

- $\tilde{V}_i(s, x) = U_i(x)$ for every $i \in N$ and for every $s, x \in X$.

Hence, $\xi$ forms a correlated equilibrium if a mediator who makes no changes in the players' payoff can run a randomization device according to $\xi$, picks a profile of pure actions $s$, and sends every player $i$ the recommendation to play $s_i$, and every player is better off obeying the recommendation if she believes that all other players obey the recommendations. It is well-known ( and it was implicitly used in the proof of Theorem 7) that if $p$ is a mixed strategy equilibrium, $\xi_p$ is a correlated equilibrium. Moreover, going over the proof of Theorem 7 reveals that the only property of the mixed-strategy equilibrium $p$ that we use is the fact that $\xi_p$ is a correlated equilibrium. Hence we get:

**Theorem 8** *Let $G(U) = (N, X, U)$ be a finite game with at least two actions for every player. Every correlated equilibrium profile $\xi$ is 0-implementable in $G(U)$ with an appropriate implementation device $I = (S, h, \tilde{V})$ in which, $S = X$ and $h = \xi$.*

Note that eventually, when the interested party implements a mixed strategy equilibrium or a correlated equilibrium with the implementation device $I$, the players are using pure strategies in the game $G(U, I)$. Because the expected value operator is linear, it can be easily seen that obeying the recommendation remains a dominant strategy for every player even if this player believes that the other players use mixed strategies in $G(U, I)$, where a mixed strategy of $i$ in $G(U, I)$ is a probability distribution $Q_i$ over the set of his pure strategies. A mixed strategy is not a natural description of behavior in $G(U, I)$. A more natural, and less computational demanding concept is the one of behavior strategy: A *behavior strategy* of $i$ in the game $G(U, I)$ is a function $c_i : S_i \to \Delta(X_i)$. Hence, a player who is using a behavioral strategy chooses a mixed strategy in a game in strategic form as a function of his signal, while a player who is using a mixed strategy in $G(U, I)$ is picking a pure strategy in $G(U, I)$ with a randomization device before he receives the signal. The sets of mixed and behavioral strategies are not technically related to each other. However, by Kuhn (1953) ( see also Hart (1992) for details), for every player $i$, for every strategy $b_i$ of $i$ and for every profile $Q_{-i} = (Q_j)_{j \neq i}$ of mixed strategies of the other players, there exists a profile $c_{-i} = (c_j)_{j \neq i}$ of behavioral strategies of the other players such that for every signal $s_i$ the expected utility of $i$ when using $b_i$ given that all other players are using $Q_{-i}$ equals his expected utility when all other players are using $c_{-i}$, and vice versa.[16]

Hence, theorems 7 and 8 remain valid in an environment that allows the utilization of either mixed or behavioral strategies in $G(U, I)$.

## 7. Conclusions, discussion and further research

One may distinguish between two main lines of research in multi-agent systems. One line of research has to do with the design of mechanisms and protocols. In this context we in fact deal with the design of games, such that when agents are assumed to behave in particular way (e.g., be law-abiding, playing in equilibrium, reinforcement learners, etc.) then a desired behavior (e.g., revenue maximization, the maximization of social surplus, etc.) will be obtained. The other line of research deals with the study of the behavior of agents in a given game. In economic theory the leading paradigm in the last decades has been that agents use equilibrium strategies. However, this paradigm implicitly assumes that agents are rational and moreover, that agents believe that the other agents are rational.[17]

In this paper we introduced an intermediate approach. The game/interaction is given, and agents are not provided with newly designed protocol. The influence on agents' behavior is only through credible promises for positive monetary transfers conditioned on the actions selected by the agents in the game. We also assume:

1. Minimal rationality: we would like to assume as little as possible on agent rationality. Indeed all that we assume is that an agent will not use a strategy if it is dominated by another strategy.

2. Minimal expenses: we assume that the interested party wishes to minimize its expenses while leading the agents to the desired behavior.

The notion of $k$-implementation captures the above basic ideas. In the paper we have provided several basic results about $k$-implementation. We have provided full characterization of $k$-implementation for the case of implementation of singletons in games with complete information. In particular we have shown a formula for computing the optimal $k$, which will yield the desired agent behavior, and fully specified a procedure for implementation. Our result is also applicable to a general class of games with infinitely many strategies. We have shown that such characterization is likely to be impossible for the implementation of sets of strategies. In particular, we have shown that the problem of whether the desired behavior can be implemented with a cost of $k$ is NP-hard. This led us to considering exact $k$-implementation that we have shown to be tractable. Exact $k$-implementation requires that every desirable strategy profile will be rational as a result of the promises for monetary transfers (in addition to the requirement that no undesirable strategy profile will remain rational). In a sense, exact implementation requires that the number of strategies that the system will "remove" will be minimal. This is consistent with basic ideas of normative systems where in order to obtain desired behavior we would like to have minimal laws that leave maximal freedom to the agents as long as they will enable to obtain the desired ("social") behavior. For a discussion of this issue see (Fitoussi & Tennenholtz, 2000).

---

16. Actually, the theorem presented by Kuhn (1953) is stronger than the one we quote here.

17. In computer science the issue of how *should* an agent choose its action , unless he has a dominant strategy, is a central one and no general satisfactory solution is known. Researchers appeal in this case to the concept of competitive analysis (Borodin & El-Yaniv, 1998). In the context of games, some promising results in this direction are presented by Tennenholtz (2002).

The extension of our results to the context of mixed strategies can be interpreted as a strong evidence for the importance of Nash equilibrium from the normative perspective, and not only as a descriptive approach that attempts to explain/predict agent behavior in economic contexts. The fact that any mixed strategy equilibrium can made into a dominant one with zero payments given that we have a credible interested party, who cannot force behaviors or punish agents, tells us that in many practical situations Nash equilibrium has a special merit also from the normative perspective.

The games of incomplete information discussed in this section are games in informational form, rather than Bayesian games. This is in the spirit of work in computer science that tries to minimize probabilistic assumptions about the economic environment, and in particular the use of those as part of a solution concept. In this context the VCG mechanisms are probably the most central and widely studied mechanisms and we turn our attention to the study of $k$-implementation in the context of these mechanisms. Indeed, as we show, unlike in the case of finite games with complete information where there is always some large $k$ that (if paid) can lead to any desired behavior, this is no longer true in games with incomplete information. The VCG mechanisms turn out to be very complex in their equilibrium analysis. As recent work has shown (Holzman et al., 2003) there are exponentially many equilibria of the VCG mechanisms that differ from truth telling. These are of special interest since these other equilibria exhibit lower communication complexity than the standard truth revealing equilibrium. Notice that these equilibria are obtained without any restriction on the possible bids by the agents (this is the so called no imposition property). The interested party has no access to the VCG protocol, and can influence it only indirectly. We have proved 0-implementation of any ex-post equilibrium of any frugal VCG mechanism, and that the frugality requirement is necessary.

There are many things left to be done. In particular, it will be interesting to further develop the study of $k$-implementation for better understanding the case of $k > 0$. For example, it may be interesting to study the effects of the cost of implementation on economic efficiency. Further study of tractable cases is also of interest. We are interested also in extending our study of $k$-implementation to games in informational form beyond the VCG mechanisms. The issue of collusion may have very interesting ramifications in the context of $k$-implementation. Collusive agreements may make benefit of the promises made by the interested party. Similarly, if failures in the system are possible, then the interested party might find itself paying its offers which he was hoping to ignore given rational behavior of the agents. We have also assumed that there is only one interested party. In a case that there are several interested parties who may wish to lead to different desired behaviors then a new strategic situation emerges. We believe that these issues are of significant importance. We hope to address some of them in future work, and that others will join us in the study of $k$-implementation and in exploring the spectrum between the system and agent perspectives in multi-agent systems.

## Acknowledgements

We thank Assaf Cohen for helpful comments. The first author thanks the Israeli Science Foundation for partial support of this research through a BIKURA grant, and the second author thanks the Israeli Science Foundations for partial support of this research by ISF individual grants. A preliminary version of this paper appears in the proceedings of the 4th ACM Conference on Electronic Commerce (EC'03).